\documentclass[
    ,final            
  ]
  {aipproc}

\layoutstyle{6x9}

\begin{document}

\title{Repulsively bound atom pairs: Overview, Simulations and Links}


\author{A. J. Daley}{
  address={Institut f\"ur Theoretische Physik, Universit\"at Innsbruck,
A-6020 Innsbruck, Austria} ,altaddress={Institut f\"ur Quantum Optik und
Quantum Information der \"Osterreichischen Akademie der Wissenschaften, A-6020
Innsbruck, Austria} }

\author{A. Kantian}{
  address={Institut f\"ur Theoretische Physik, Universit\"at Innsbruck,
A-6020 Innsbruck, Austria} ,altaddress={Institut f\"ur Quantum Optik und
Quantum Information der \"Osterreichischen Akademie der Wissenschaften, A-6020
Innsbruck, Austria} }

\author{H. P. B\"uchler}{
  address={Institut f\"ur Theoretische Physik, Universit\"at Innsbruck,
A-6020 Innsbruck, Austria} ,altaddress={Institut f\"ur Quantum Optik und
Quantum Information der \"Osterreichischen Akademie der Wissenschaften, A-6020
Innsbruck, Austria} }

\author{P. Zoller}{
  address={Institut f\"ur Theoretische Physik, Universit\"at Innsbruck,
A-6020 Innsbruck, Austria} ,altaddress={Institut f\"ur Quantum Optik und
Quantum Information der \"Osterreichischen Akademie der Wissenschaften, A-6020
Innsbruck, Austria} }

\author{K. Winkler}{
  address={Institut f\"ur Experimentalphysik, Universit\"at Innsbruck,
A-6020 Innsbruck, Austria} }

\author{G. Thalhammer}{
  address={Institut f\"ur Experimentalphysik, Universit\"at Innsbruck,
A-6020 Innsbruck, Austria} }

\author{F. Lang}{
  address={Institut f\"ur Experimentalphysik, Universit\"at Innsbruck,
A-6020 Innsbruck, Austria} }

\author{R. Grimm}{
  address={Institut f\"ur Experimentalphysik, Universit\"at Innsbruck,
A-6020 Innsbruck, Austria} ,altaddress={Institut f\"ur Quantum Optik und
Quantum Information der \"Osterreichischen Akademie der Wissenschaften, A-6020
Innsbruck, Austria} }

\author{J. Hecker
Denschlag}{
  address={Institut f\"ur Experimentalphysik, Universit\"at Innsbruck,
A-6020 Innsbruck, Austria} }

\begin{abstract}
We review the basic physics of repulsively bound atom pairs in an optical
lattice, which were recently observed in the laboratory \cite{Win2006},
including the theory and the experimental implementation. We also briefly
discuss related many-body numerical simulations, in which time-dependent
Density Matrix Renormalisation Group (DMRG) methods are used to model the
many-body physics of a collection of interacting pairs, and give a comparison
of the single-particle quasimomentum distribution measured in the experiment
and results from these simulations. We then give a short discussion of how
these repulsively bound pairs relate to bound states in some other physical
systems.
\end{abstract}

\maketitle


\subsection{Introduction}

 Stable bound states, in which the composite
object has a lower energy than the separated constituents, give rise to much
diversity and complexity in many physical systems. Well-known examples ranging
from chemically bound atomic molecules to excitons in solid state physics rely
on attractive interactions to give rise to bound objects. The converse,
particles bound by a repulsive interaction, is impossible in free space because
interaction energy can be freely converted to kinetic energy of the constituent
atoms. However, by placing particles on a lattice, kinetic energy is restricted
to fall within the Bloch bands, and repulsively interacting atoms cannot always
move apart and convert their interaction energy to kinetic energy. Recently we
have reported on the first clear observation of such states, in the form of
repulsively bound pairs of atoms in an optical lattice \cite{Win2006}.

The stability of these pairs relies on the weak coupling of atoms in optical
lattices to dissipative processes, which would otherwise lead to rapid
relaxation of the system to its ground state (as is typically seen, e.g., in
the context of solid state lattices). In this article we give an overview of
repulsively bound atom pairs, beginning with a discussion of a single pair, and
proceeding with a discussion of the experimental implementation, and many-body
numerical simulations used to analyse a system of many interacting pairs. We
then comment on analogies between these composite objects and bound states
found in other physical systems.

\subsection{Repulsively bound atom pairs in an optical lattice}

The existence of repulsively bound atom pairs is predicted by the Bose-Hubbard
model \cite{Fis89}, which describes well the dynamics of ultracold atoms loaded
into the lowest band of a sufficiently deep optical lattice
\cite{hubbardtoolbox}. The corresponding Hamiltonian is
\begin{equation}
\hat{H} =-J\sum_{\langle i,j\rangle }{\hat{b}}_{i}^{\dag }{\hat{b}}_{j}
+\frac{U}{2}\sum_{i}{\hat{n}}_{i}\left( {\hat{n}}_{i} -1\right), \label{BH}
\end{equation}%
where ${\hat{b}}_{i}$ (${\hat{b}}_{i}^{\dag }$) are destruction (creation)
operators for the bosonic atoms at site $i$ of the lattice, and ${\hat{n}}_{i}
= {\hat{b}}_{i}^{\dag }{\hat{b}}_{i}$ is the corresponding number operator.
$J/\hbar$ denotes the nearest neighbor tunnelling rate, and $U$ the on-site
collisional energy shift. The relative value of $U$ and $J$ can be adjusted by
varying the depth of the lattice $V_0$.

In the limit of $U/J\rightarrow \infty$, the repulsively bound pair can be seen
as an object where two atoms are located on the same lattice site. Due to the
interaction between atoms, this state has an energy offset of $U$ compared with
states where atoms are present on different lattice sites. The stability of the
pair can then be understood by energy conservation arguments: Two separated
atoms moving in the lowest Bloch band of a lattice can have a maximum combined
kinetic energy of $8J$ (in 1D). Thus, the atoms cannot separate and convert
their interaction energy to kinetic energy (see Figure 1).

\begin{figure}
  \includegraphics[width=0.7\textwidth]{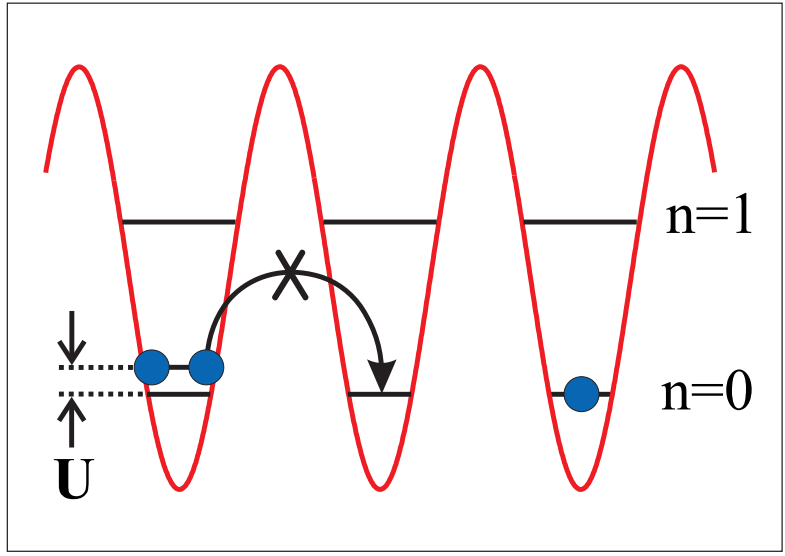}
  \caption{A state with two atoms located on the same site of an optical lattice has an energy offset $\approx U$
  with respect to states where the atoms are separated. Breaking up of
the pair is suppressed due to the lattice band structure and energy
conservation, so that the pair remains bound as a composite object, which can
tunnel through the lattice. In the figure, $n=0$ denotes the lowest Bloch band
and $n=1$ the first excited band.}
\end{figure}

More generally, repulsively bound pairs arise from the eigenstates of the
Bose-Hubbard model with two atoms present on the lattice. Denoting the
primitive lattice vectors in each of the $d$ dimensions by $\mathbf{e}_{i}$, we
can write the position of the two atoms by $\mathbf{x}=\sum_{i=1}^dx_{i}
\mathbf{e}_{i}$ and $\mathbf{y}=\sum_{i=1}^dy_{i} \mathbf{e}_{i}$, where
$x_{i}, y_i$ are integers, and we can write the two atom wave function in the form $%
\Psi (\mathbf{x},\mathbf{y})$. The related Schr\"odinger equation from the
Bose-Hubbard model then takes the form
\begin{equation}
\left[ -J \left(\tilde{\Delta}^{0}_{{\bf x}} + \tilde{\Delta}_{{\bf y}}^{0}
\right) + U \delta_{{\bf x},{\bf y}}\right] \Psi({\bf x},{\bf y}) = E \:
\Psi({\bf x},{\bf y}), \label{schroedinger}
\end{equation}
where the operator
\begin{equation}
\tilde{\Delta}^{{\bf K}}_{{\bf x}} \Psi({\bf x})\!=\! \sum^{d}_{i=1}
\cos\left({\bf K}{\bf e}_{i}/2\right)\left[ \Psi({\bf x\!+\!e}_{i}) \!+\!
\Psi({\bf x\!-\!e}_{i}) - 2 \Psi({\bf x})\right]
\end{equation}
denotes the discrete lattice Laplacian on a cubic lattice. Writing the
wavefunction in relative and centre of mass coordinates $\Psi({\bf x}, {\bf y})
= \exp(i {\bf K} {\bf R}) \psi_{\bf K}({\bf r})$, the Schr\"odinger equation
can be reduced to a single particle problem in the relative coordinate
\begin{equation}
  \left[ -2 J \tilde{\Delta}^{\bf K}_{\bf r}  + E_{\bf K}
  + U \delta_{{\bf r},0}\right]
\psi_{\bf K}({\bf r}) = E \psi_{\bf K}({\bf r}) \label{singleparticleSE}
\end{equation}
where $E_{\bf K}= 4 J\sum_{i=1} \left[1\!-\!\cos ({\bf K}{\bf e}_{i}/2)
\right]$ is the kinetic energy of the center of mass motion.

The short range interaction potential makes it possible to resum the
perturbation expansion for the associated Lippman-Schwinger equation, and we
obtain the scattering states
\begin{equation}
 \psi_{E}({\bf r}) = \exp( i {\bf k} {\bf r}) - 8 \pi J f_E({\bf K}) G_{{\bf K}}(E, {\bf r}) \label{scatteringstates}
\end{equation}
with scattering amplitude
\begin{equation}
f_E({\bf K})= -\frac{1}{4\pi}\frac{U/(2J)  }{1-G_{\bf K}(E,0) U}
\end{equation}
with total energy $E= \epsilon_{{\bf k},{\bf K}} + E_{{\bf K}}$, and
$\epsilon_{{\bf k},{\bf K}}= 4 J \sum_{i=1} \cos({\bf K}{\bf e}_{i}
/2)\left[1-\cos({\bf k}{\bf e}_{i})\right]$. Furthermore, $G_{{\bf K}}(E,{\bf
r})$ denotes the Greens function of the non-interacting problem, which in
Fourier space takes the form $\tilde G_{\bf K}(E,{\bf k}) =1/(E- \epsilon_{{\bf
k},{\bf K}} - E_{\bf K} + i \eta)$. The scattering states $\psi_{E}({\bf r})$
correspond to two free atoms moving on the lattice and undergoing scattering
processes.

 In addition, the pole in the scattering
amplitude indicates the presence of an additional bound state for each value of
$\bf K$, which corresponds to the repulsively bound pair. The energy $E_{\rm
bs}$ of the bound states is determined by $1= U G_{\bf K}(E_{\rm bs},0)$ while
the bound state wave function takes the form $\psi^{\rm bs}_{\bf K}({\bf r}) =
c \: G_{\bf K}(E_{\rm bs},{\bf r})$ with $c$ a normalization factor.  Note that
in one dimension such bound states exists for arbitrary repulsive interaction,
but for a three-dimensional lattice such bound states, and therefore
repulsively bound pairs, appear only for a repulsive interaction above a
critical value $U
> U_{\rm crit} \approx 8 J$ (for $K=0$). These states have a square-integrable
relative wavefunction $\psi_{\bf K}({\bf r})$, as shown for two different
values of $U/J$ in Figure 2. For a deep lattice, i.e. $U/J \gg 1$, bound pairs
essentially consist of two atoms occupying the same site, whereas for small
$U/J$, the pair is delocalized over several lattice sites. A main feature of
the repulsive pair wavefunction is its oscillating character: the wavefunction
amplitude alternates sign from one site to the next. In quasimomentum space
this corresponds to a wavefunction which is peaked at the edges of the first
Brillouin zone, as is shown in Figure 3.

\begin{figure}
  \includegraphics[width=0.9\textwidth]{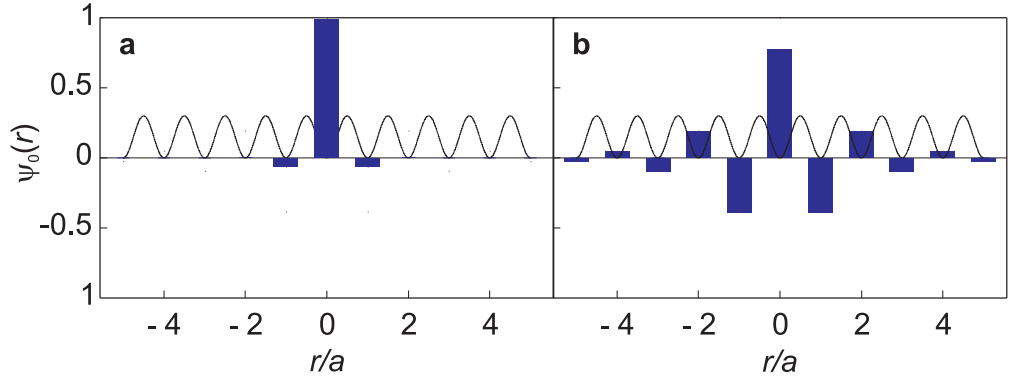}
  \caption{Relative wavefunctions $\psi_{\bf K}({\bf r})$ for repulsively bound pairs
($a_s = 100 a_0$) in 1D with $K=0$, for (a) $U/J=30$ $(V_0 = 10 E_r)$ and (b)
$U/J=3$ $(V_0 = 3 E_r)$, where $E_r$ is the recoil energy.
($E_r=2\pi^2\hbar^2/m\lambda^2$, where $m$ is the mass of the atoms and
$\lambda$ is the twice the lattice period, $a$.)}
\end{figure}

\begin{figure}
  \includegraphics[width=0.9\textwidth]{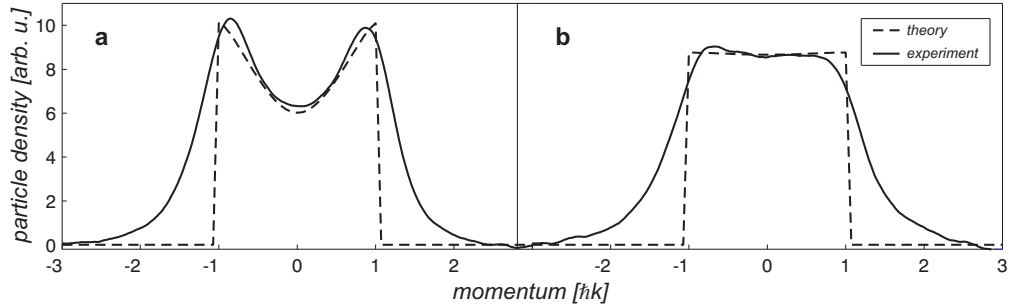}
  \caption{Single particle quasimomentum distributions for repulsively bound pairs
 in 1D from experiment and numerical simulations (see text) for (a) $V_0=5E_R$ and (b) $V_0=20E_R$. The
 density values have been scaled to facilitate comparison between experimental and theoretical results. These results agree well, up to
experimental artifacts related to repulsion between atoms during expansion
(before imaging) and also relatively long imaging times (many photons are
scattered from each atom, which performs a random walk). This leads to smearing
out of the sharp structure at the edge of the Brillouin zone.}
\end{figure}

When many repulsively bound pairs exist, they will interact with one another as
described by the Bose-Hubbard model. This many-body behaviour can be computed
numerically as described below.

\subsection{Experimental realization of repulsively bound atom pairs}
We experimentally create repulsively bound atom pairs from a sample of
ultracold $^{87}$Rb atoms in a cubic 3D optical lattice. About $2 \times 10^4$
atom pairs are initially prepared in a deep lattice of 35$E_r$ depth, with each
site of the lattice either doubly-occupied or unoccupied. By adiabatically
ramping down the lattice depth afterwards the initially localized pair
wavefunctions become delocalized (see Fig. 2). The initial preparation is
carried out in several steps as described in the following. In the beginning, a
Bose Einstein condensate of $^{87}$Rb atoms is carefully loaded into the
vibrational ground state of the optical lattice, such that many lattice sites
are occupied with two atoms. Besides the doubly occupied sites, there are also
sites which are occupied by single atoms or more than two atoms. In order to
remove atoms from these sites we use a purification scheme which involves the
use of a Feshbach resonance and a combined pulse of laser and radio-frequency
(rf) radiation \cite{Tha06}. The laser and rf pulse resonantly blows atoms out
of the lattice, whereas the Feshbach resonance serves to protect (shelve) the
pairs temporarily from this pulse by converting pairs into Feshbach molecules
and then back into atoms. Besides lifetime measurements, we have been able to
experimentally map out the single particle momentum distribution (see Fig. 3)
and to measure their binding energy. The properties and the dynamics of the
pairs can be controlled by tuning the atom-atom interaction with the help of a
Feshbach resonance at 1007G and by controlling the depth of the optical lattice
and particle density. Consistent with our theoretical analysis, the repulsively
bound pairs exhibit long lifetimes of hundreds of milliseconds, even under
collisions with one another.

\subsection{Many-Body simulations}

Many-body numerical simulations for a gas of repulsively bound pairs are
performed using time-dependent DMRG methods \cite{td-dmrg}. These methods allow
for ground state calculation and time-dependent calculation of the dynamics of
atoms for a variety of 1D situations, including many lattice and spin models.
The basic algorithm provides near-exact integration of a many-body
Schr\"odinger equation, with the Hilbert space being adaptively decimated. This
works provided that the state of the system is always able to be efficiently
represented as a matrix product state \cite{dmrgreview}. As a result, it is
possible to compare the dynamics of a gas of interacting repulsively bound
pairs in a 1D lattice with experimental data. For example, we can simulate a 1D
Bose-Hubbard model with time dependent parameters, beginning with an initial
state corresponding to a distribution of atoms situated in doubly occupied
lattice sites. We compute the corresponding dynamics as the lattice depth is
decreased by decreasing $U$ and increasing $J$. These many-body simulations
account for interactions between bound pairs, and let us compute final momentum
distributions that agree well with the experimental results. We can also use
these simulations to model lattice modulation spectroscopy of atoms in optical
lattices. In figure 3 we show a comparison of quasimomentum distributions from
the experiment and from many-body simulations.

\subsection{Analogy to Other Bound States}

Although no stable repulsively bound pairs have previously been observed, they
have an interesting relationship to many bound states in other physical
systems. For example, resonance behaviour based on similar pairing of Fermions
of different spin in the Hubbard model was first discussed by Yang
\cite{Yan1989}, and plays an important role in SO(5) theories of
superconductivity \cite{Han2004}. There are several examples of many-body bound
states that can occur for repulsive as well as attractive interactions, such as
the resonances discussed in the context of the Hubbard model by Demler et al.
\cite{Dem1995}. Such resonance behaviour is common in many-body physics,
although states of this type are normally very short-lived. Optical lattice
experiments will now provide an opportunity to prepare and investigate stable
versions of such states, which until now have only appeared virtually as part
of complex processes.

The stability and many-body physics of repulsively bound pairs is perhaps most
closely associated with that of excitons, which are bound pairs of a particle
in the conduction band and a hole in the valence band of a periodic system
\cite{Moskalenko}. These bind to form a composite boson, a gas of which can, in
principle, Bose-condense. Excitons are excited states of the many-body system,
but are bound by an attractive interaction between the particle and hole that
form the pair. They are also discussed in the specific context of fermionic
systems. However, a single exciton on a lattice could have a description very
similar to that of a single repulsively bound pair, and could be realised and
probed in optical lattices experiments \cite{Kan06}.

Repulsively bound atom pairs in an optical lattice are also reminiscent of
photons being trapped by impurities in photonic crystals
\cite{photonicCrystal}, which consist of transparent material with periodically
changing index of refraction. An impurity in that crystal in form of a local
region of index of refraction can then give rise to a localized field
eigenmode. In an analogous sense, each atom in a repulsively bound pair could
be as an impurity that ``traps'' the other atom.

An analogy can also be drawn between repulsively bound atom pairs and gap
solitons, especially as found in atomic gases
\cite{Lou2003,Efr2003,Eie2004,Ahu04}. Solitons are normally a non-linear wave
phenomenon, and in this sense have a very different behaviour to repulsively
bound pairs, which exhibit properties characteristic of many-body quantum
systems. However, there has been increasing recent interest in discussing the
limit of solitons in atomic systems where very few atoms are present, giving
rise to objects that are often referred to as quantum solitons
\cite{quantumsolitons}. These are N-body bound states in 1D, and thus a 2-atom
bright quantum soliton is a bound state of two atoms moving in 1D. In this
sense, the solution for a single repulsively bound pair in 1D is related to a
single quantum soliton on a lattice.



\subsection{Conclusion}

In summary, a metastable bound state that arises from repulsion between the
constituents and the lattice band structure has been demonstrated in the
laboratory. This state exemplifies in a new way the strong correspondence
between the optical lattice physics of ultracold atoms and the Hubbard model, a
connection which has particular importance for applications of these cold atom
systems to more general simulation of condensed matter models, to quantum
computing. The existence of such metastable bound objects will be ubiquitous in
cold atoms lattice physics, giving rise to new potential composite objects also
in Fermions or in systems with mixed Bose-Fermi statistics. These states could
also be formed with more than two particles, or as bound states of existing
composite particles. Repulsively bound pairs have no counterpart in condensed
matter physics due to the strong inelastic decay channels observed in solid
state lattices, and could be a building block of yet unstudied quantum many
body states or phases.

\begin{theacknowledgments}
We would like to thank Eugene Demler for interesting discussions. We
acknowledge support from the Austrian Science Fund (FWF) within the
Spezialforschungsbereich 15, from the European Union within the OLAQUI and
SCALA networks, from the TMR network "Cold Molecules", and the Tiroler
Zukunftsstiftung.
\end{theacknowledgments}

\end{document}